\documentclass[letterpaper, 10 pt, conference]{ieeeconf}  
\IEEEoverridecommandlockouts                              
\usepackage{cite}
\usepackage{amsmath,amssymb,amsfonts}
\usepackage{graphicx}
\usepackage{textcomp}
\usepackage{xcolor}
\usepackage{booktabs}
\usepackage{multirow}
\usepackage{subcaption}

\title{\LARGE \bf
MA-IDS: Multi-Agent RAG Framework for IoT Network Intrusion Detection with an Experience Library%
\author{Md Shamimul Islam $^{1}$, Luis G. Jaimes$^{2}$ and Ayesha S. Dina$^{3}$
\thanks{$^{1}$Md Shamimul Islam is with the Department of Computer Science, Florida Polytechnic University,
        Lakeland, Florida
        {\tt\small mislam3051@floridapoly.edu}}%
\thanks{$^{2}$Luis G. Jaimes is with the Department of Computer Science, Florida Polytechnic University,
        Lakeland, Florida
        {\tt\small ljaimes@floridapoly.edu}}%
\thanks{$^{3}$Ayesha S. Dina is with the Department of Computer Science, Florida Polytechnic University,
        Lakeland, Florida
        {\tt\small adina@floridapoly.edu}}%
\thanks{This work has been submitted to the IEEE for possible publication. 
Copyright may be transferred without notice, after which this version may no longer be accessible.}
}}
\begin{document}
\maketitle
\thispagestyle{empty}
\pagestyle{empty}


\begin{abstract}
    Network Intrusion Detection Systems (NIDS) face important limitations. Signature-based methods are effective for known attack patterns, but they struggle to detect zero-day attacks and often miss modified variants of previously known attacks, while many machine learning approaches offer limited interpretability. These challenges become even more severe in IoT environments because of resource constraints and heterogeneous protocols. To address these issues, we propose MA-IDS, a Multi-Agent Intrusion Detection System that combines Large Language Models (LLMs) with Retrieval Augmented Generation (RAG) for reasoning-driven intrusion detection. The proposed framework grounds LLM reasoning through a persistent, self-building Experience Library. Two specialized agents collaborate through a FAISS-based vector database: a Traffic Classification Agent that retrieves past error rules before each inference, and an Error Analysis Agent that converts misclassifications into human-readable detection rules stored for future retrieval, enabling continual learning through external knowledge 
    accumulation, without modifying the underlying language model. Evaluated on NF-BoT-IoT and NF-ToN-IoT benchmark datasets, MA-IDS achieves Macro F1-Scores of 89.75\% and 85.22\%, improving over zero-shot baselines of 17\% and 4.96\% by more than 72 and 80 percentage points. These results are competitive with SVM while providing rule-level explanations for every classification decision, demonstrating that retrieval-augmented reasoning offers a principled path toward explainable, self-improving intrusion detection for IoT networks.
    
\end{abstract}

\section{INTRODUCTION}
The proliferation of Internet of Things (IoT) devices has exponentially expanded the network attack surface, precipitating a rise in sophisticated, high-volume cyber threats. Traditional Network Intrusion Detection Systems (NIDS) primarily rely on signature-based methods, which are inherently reactive; they struggle to identify zero-day exploits or polymorphic variations of known attacks, leading to high false-negative rates~\cite{b1,b2}. 

To overcome these limitations, machine learning (ML) and deep learning (DL) approaches have been widely adopted for their ability to model complex patterns in high-dimensional data. However, these models typically operate as ``black boxes,'' providing high classification accuracy at the expense of interpretability. This is a critical shortcoming in security-critical environments where forensic justification is required for incident response. Furthermore, their reliance on static training distributions makes them brittle against evolving threats, often necessitating computationally expensive retraining to maintain efficacy~\cite{b2,b4}.

Large Language Models (LLMs) offer a paradigm shift by enabling semantic reasoning, few-shot generalization, and human-readable explanations. Yet, their direct application to NIDS is hindered by a ``domain gap'': pre-trained LLMs struggle to interpret numerical NetFlow telemetry, which lacks the natural language structure they were trained on. Our empirical benchmarks highlight this failure: a zero-shot GPT-4o baseline achieves macro F1-scores of only 17\% on NF-BoT-IoT and a negligible 4.96\% on NF-ToN-IoT. These results confirm that without domain-specific grounding, LLMs provide unstable and often misleading threat assessments~\cite{b4, zhang2024large}.

To bridge this gap without the prohibitive overhead of continuous fine-tuning, we propose MA-IDS, a Multi-Agent Intrusion Detection System that transforms classification errors into a persistent, self-evolving Experience Library. Unlike traditional supervised models that suffer from ``silent failure,'' MA-IDS employs a closed-loop architecture featuring two specialized GPT-4o agents: (1) an Error Analysis Agent that identifies discriminative features in misclassified flows to formulate structured, semantic rules; and (2) a Traffic Classification Agent that utilizes Retrieval-Augmented Generation (RAG) to query a FAISS-based vector database for relevant past experiences during inference.

Evaluated on the NF-BoT-IoT and NF-ToN-IoT datasets, MA-IDS achieves macro F1-scores of 89.75\% and 85.22\%, respectively, which constitutes a performance leap of up to 80 percentage points over zero-shot baselines. While supervised classifiers like SVM may achieve higher peak accuracy on static benchmarks, MA-IDS offers a critical trade-off: semantic resilience. By providing rule-level forensic justifications for every decision, it shifts NIDS from a pattern-matching paradigm to an adaptive reasoning framework capable of continuous self-improvement without parameter modification.


The remainder of this paper is organized as follows. Section II reviews the evolution of NIDS from traditional ML to agentic AI. Section III details the MA-IDS architecture, including its multi-agent workflow and Experience Library. Section IV presents the experimental setup, performance results, and comparative analysis. Finally, Section V summarizes the key findings and outlines future research directions.

\begin{figure*}[t]
\centering
\includegraphics[width=0.76\textwidth]{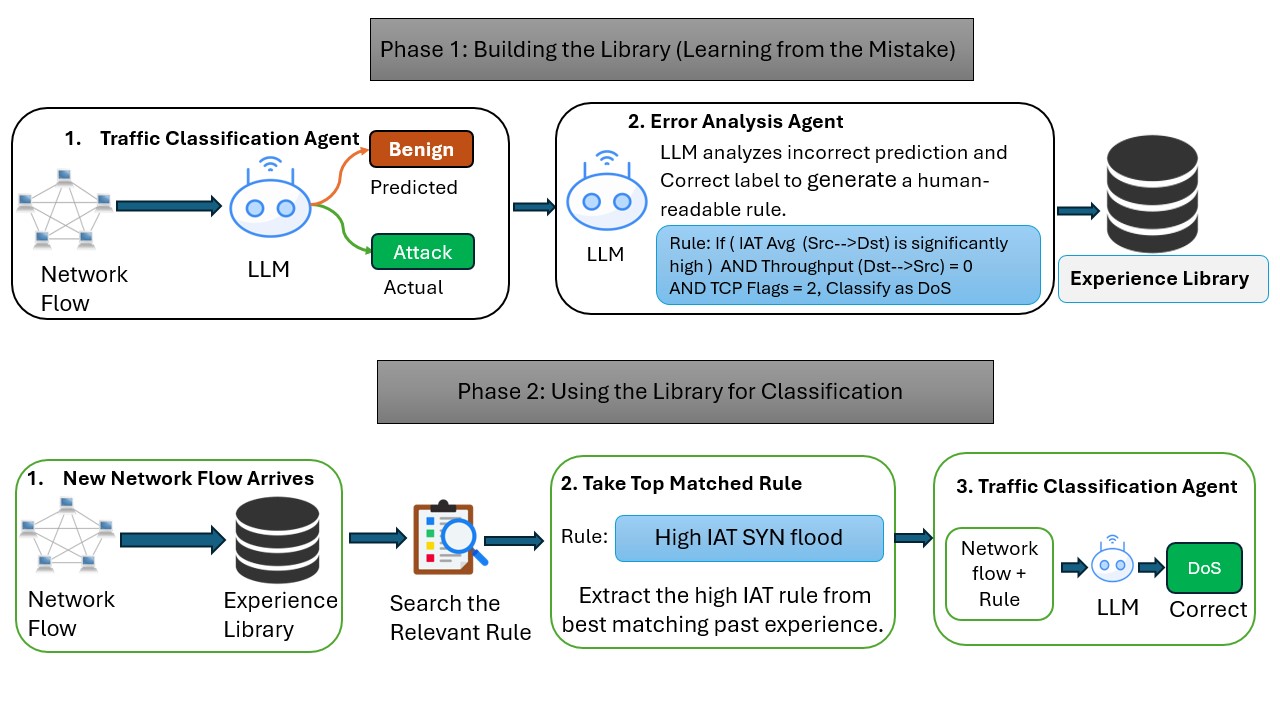}
\caption{Dual-Phase Agentic Workflow for MA-IDS. Phase 1 (Top) illustrates the offline library-building loop, where an Error Analysis Agent evaluates misclassifications to generate human-readable rules for the Experience Library. Phase 2 (Bottom) depicts the real-time classification process, utilizing RAG to inject past experiences into the LLM’s context for informed decision-making.}
\label{fig:wide_architecture}
\end{figure*}

\section{RELATED WORK}

\textbf{Conventional IDS.} Foundational research established two primary paradigms: signature-based systems (SIDS), which identify known threats through predefined patterns~\cite{ioulianou2018signature, nagaraju2021high, alyousef2019dynamically, denning1987intrusion}, and anomaly-based systems (AIDS), which flag deviations from established normal behavior~\cite{khraisat2018anomaly, jia2021network, axelsson1998research, bace2001intrusion}. While deployment guidelines for network and host-based configurations are well-documented~\cite{scarfone2007guide, khraisat2019survey}, these systems face significant maintenance burdens. Knowledge-driven variants employing formal description languages~\cite{studnia2018language} or rule-based expert systems~\cite{kim2014novel} often struggle in dynamic network environments. Critically, SIDS are ineffective against zero-day exploits~\cite{ullah2019cyber}, and AIDS require frequent manual recalibration to maintain accuracy. MA-IDS addresses these gaps by bypassing static signatures in favor of autonomous, error-driven rule induction that eliminates the need for manual profile updates.

\textbf{Machine Learning-Based IDS.} A variety of machine learning (ML) algorithms, including multilayer perceptrons~\cite{bajaj2013dimension}, k-nearest neighbors~\cite{li2014new}, decision trees~\cite{guezzaz2021reliable}, and SVMs~\cite{mohammadi2021comprehensive}, have been evaluated for anomaly-oriented detection. In IoT environments, ensemble methods like Random Forest and AdaBoost have been extensively analyzed~\cite{verma2020machine}, alongside lightweight models for resource-constrained deployments~\cite{roy2022lightweight} and quantum annealing for feature selection~\cite{davis2024quantum}. Despite their classification efficacy, these models function as ``black boxes'' and require full retraining to incorporate new attack vectors. MA-IDS maintains competitive performance while providing human-readable semantic explanations and adapting to new threats without model retraining.

\textbf{Deep Learning-Based IDS.} Deep learning (DL) architectures, such as CNNs, LSTMs, and autoencoders, have become central to modern NIDS research~\cite{xu2025deep, survey2024applications, intrusion2024detection, ferrag2019deep}. Recent efforts have focused on IoT-specific threat detection~\cite{novel2023deep}, hybrid CNN-BiLSTM models~\cite{hybrid2024deep}, and explainable DL frameworks for Industrial IoT~\cite{explainable2023deep, multi2024deep}. However, DL models remain fundamentally static post-training and demand substantial labeled data for retraining. In contrast, MA-IDS generates interpretable rules during inference, operates effectively without massive labeled datasets, and provides real-time decision-time reasoning.


\textbf{LLM and Agentic AI-Based IDS.} The application of Large Language Models (LLMs) to network intrusion detection has gained increasing attention, with techniques such as in-context learning improving detection performance~\cite{zhang2024large}. TrafficGPT~\cite{b2} leverages decoder-only architectures for open-set classification, while NetMoniAI~\cite{b3} explores hierarchical multi-agent coordination for distributed monitoring. IDS-Agent~\cite{b1} further incorporates reasoning–action pipelines and long-term memory to integrate outputs from multiple ML models. 

However, existing approaches lack mechanisms for systematic self-correction based on deployment feedback. MA-IDS addresses this limitation by introducing an Error Analysis Agent that converts misclassifications into verifiable rules stored in a persistent Experience Library. This enables a self-improving feedback loop, reducing reliance on ungrounded zero-shot reasoning and enhancing contextual decision-making.

\section{METHODOLOGY}

The MA-IDS framework is designed to bridge the gap between high-dimensional network telemetry and the semantic reasoning capabilities of Large Language Models (LLMs). As illustrated in Figure 1, the system integrates two specialized agents that interact through a shared, FAISS-based Experience Library. This closed-loop architecture enables the system to transform transient classification errors into persistent, human-readable knowledge, ensuring that the detection engine becomes progressively more accurate while avoiding the computational overhead of fine-tuning cycles.

\subsection{Multi-Agent Architecture}
The core architecture of MA-IDS follows a reasoning-followed-by-action pipeline. This design decouples real-time classification logic from diagnostic learning logic. Let $\mathbf{x} \in \mathbb{R}^{14}$ represent the feature vector of an incoming network flow, $\mathcal{C}$ denote the set of target traffic classes, and $\mathcal{L}_t$ represent the state of the Experience Library at time step $t$. The final classification decision $\hat{y}$ is formally defined as:
\begin{equation}
\hat{y} = \mathcal{A}_{\text{cls}} \Bigl( \mathbf{x} \;\Vert\; \mathcal{R}(\mathbf{x}; \mathcal{L}_t) \Bigr),
\label{eq:classification}
\end{equation}
where $\Vert$ denotes prompt concatenation and $\mathcal{R}(\mathbf{x}; \mathcal{L}_t)$ is the retrieval function that provides the agent with contextually relevant past experiences.

\subsubsection{Traffic Classification Agent}
The Traffic Classification Agent serves as the primary inference engine operating in the online inference phase. When a new network flow arrives, the agent utilizes Retrieval-Augmented Generation (RAG) to identify the most similar historical error stored in the library. A 384-dimensional embedding is generated for $\mathbf{x}$ using the all-MiniLM-L6-v2 model, which is then used to query the FAISS index via cosine similarity:
\begin{equation}
\mathcal{R}(\mathbf{x}; \mathcal{L}_t) =
\begin{cases}
\rho_i & \text{if } \cos\!\bigl(\operatorname{emb}(\mathbf{x}),\, \operatorname{emb}(\mathbf{z}_i)\bigr) \geq \tau, \\
\epsilon & \text{otherwise,}
\end{cases}
\label{eq:retrieval}
\end{equation}
where $\tau$ is the similarity threshold and $\rho_i$ is the retrieved semantic rule. By injecting this rule into the structured prompt $p(\mathbf{x}, r)$, the agent gains a grounded memory of past classification boundaries, significantly reducing the likelihood of repeating previous errors.

\subsubsection{Error Analysis and Rule Induction Agent}
The Error Analysis Agent constitutes the system's learning mechanism, operating in the offline refinement phase. It is activated exclusively upon a misclassification ($\hat{y} \neq y$), where $y$ is the ground-truth label. The agent performs a diagnostic ``compare-and-isolate'' reasoning chain by examining the misclassified flow $\mathbf{x}$ alongside the correct label $y$ and the erroneous prediction $\hat{y}$. The agent then distills this analysis into a concise, human-readable rule $\rho$:
\begin{equation}
\rho = \mathcal{A}_{\text{ind}} \Bigl( \mathbf{x},\ \hat{y},\ y,\ \mathcal{R}(\mathbf{x}; \mathcal{L}_t) \Bigr).
\label{eq:rule_induction}
\end{equation}
For example, the agent may identify that a high average inter-arrival time (IAT) is a primary indicator of a specific attack type that was previously mislabeled. This induced rule is committed to the Experience Library, effectively teaching the system the discriminative features between traffic classes.

\subsubsection{Experience Library and Vector Database}
The Experience Library serves as the long-term memory of MA-IDS, implemented as a persistent local FAISS vector database. Unlike traditional machine learning models that store knowledge in static mathematical weights, MA-IDS externalizes knowledge into a retrievable library, allowing the system to scale its expertise indefinitely. Each entry in $\mathcal{L}$ consists of a vector key (the flow embedding) and a metadata payload containing the semantic rule. The update process is defined as:
\begin{equation}
\mathcal{L}_{t+1} = \mathcal{L}_t \cup \Bigl\{ \bigl( \operatorname{emb}(\mathbf{x}),\ \rho,\ (\hat{y}, y) \bigr) \Bigr\}.
\label{eq:memory_update}
\end{equation}
This design choice ensures that learning is non-destructive, protecting the system from the ``catastrophic forgetting'' typically associated with deep learning fine-tuning.

\subsection{Operational Workflow}
As depicted in the Phase 1 and Phase 2 loops of Figure \ref{fig:wide_architecture}, the MA-IDS workflow operates in two interleaved stages that separate knowledge acquisition from inference.

In Phase 1 (Offline Experience Creation), the system identifies classification failures and updates the library. This process functions as an automated forensic analysis that generates a growing repository of ``hard-negative'' examples derived from the system's own errors. 

In Phase 2 (Online Classification), the system uses the library to ground its real-time decisions. Upon flow arrival, the Classification Agent performs high-speed retrieval from the Experience Library. By combining raw numerical features with high-level semantic rules, the agent performs context-aware reasoning that addresses the limitations of standard zero-shot LLM prompts.







\section{EXPERIMENTAL RESULTS}

\subsection{Experimental Setup}

The experimental evaluation was conducted on a workstation with an AMD Ryzen 5 2500U processor and 16GB of RAM. Flow embeddings were generated locally using the HuggingFace \textit{all-MiniLM-L6-v2} encoder, selected for its balance between semantic representation quality and low-latency CPU inference. The reasoning and decision-making components were implemented using OpenAI GPT-4o accessed via API. To ensure deterministic and reproducible results, the sampling temperature for both agents was fixed at 0.0.

\begin{figure*}[t]
\centering
\begin{minipage}[t]{0.5\textwidth}
    \centering
    \includegraphics[width=\linewidth, height=4.8cm, keepaspectratio]{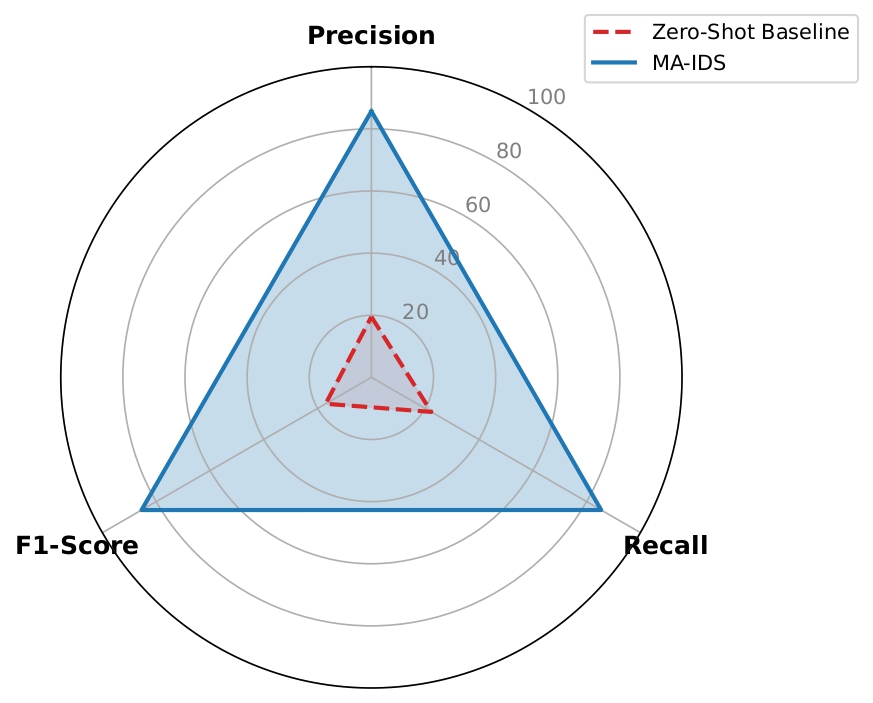}
    \subcaption{NF-BoT-IoT}
    \label{fig:radar-bot}
\end{minipage}%
\hfill
\begin{minipage}[t]{0.5\textwidth}
    \centering
    \includegraphics[width=\linewidth, height=4.8cm, keepaspectratio]{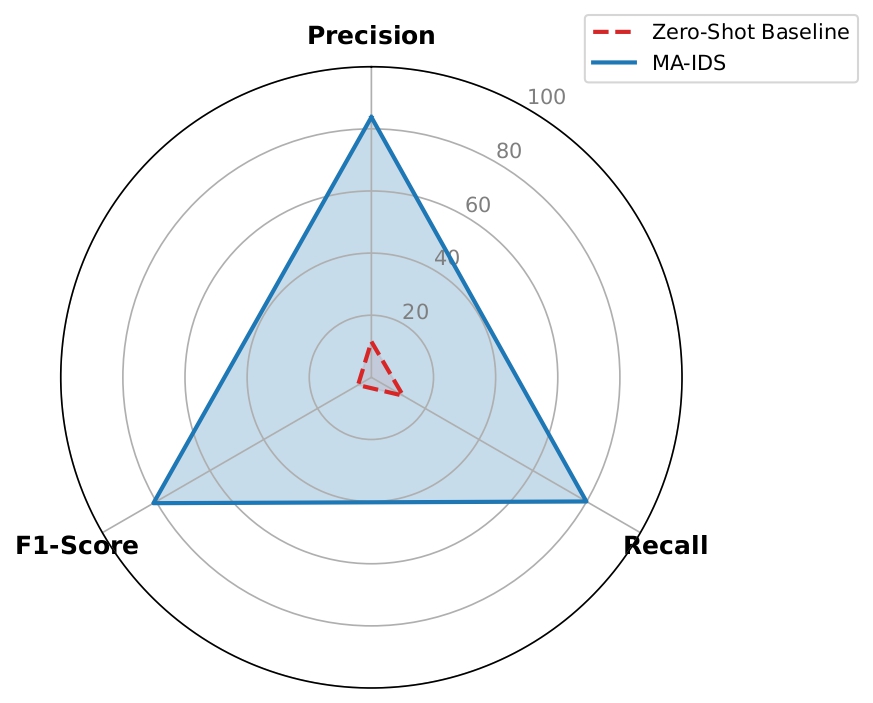}
    \subcaption{NF-ToN-IoT}
    \label{fig:radar-ton}
\end{minipage}

\caption{Macro-averaged Precision, Recall, and F1-Score for the
Zero-Shot Baseline and MA-IDS on (a) NF-BoT-IoT and (b) NF-ToN-IoT.
The near-collapsed baseline polygon versus the substantially expanded
MA-IDS polygon illustrates the critical role of Experience Library
retrieval in enabling reliable classification across all three metrics.}

\label{fig:radar-ablation}
\end{figure*}

\begin{table*}[t]
\centering
\caption{Performance comparison of MA-IDS against traditional ML baselines
and the Zero-Shot GPT-4o baseline on NF-BoT-IoT and NF-ToN-IoT test
samples. MA-IDS results reflect the evaluation stage with the Experience
Library fixed and the Error Analysis Agent disabled.}
\label{tab:comprehensive-comparison}
\small
\begin{tabular}{llcccc}
\hline
\textbf{Dataset} & \textbf{Method} & \textbf{Accuracy (\%)} & \textbf{Precision (\%)} & \textbf{Recall (\%)} & \textbf{F1-Score (\%)} \\
\hline
\multicolumn{6}{l}{\textit{NF-BoT-IoT}} \\
\hline
& AdaBoost            & 70.41 & 67.58 & 70.00 & 88.60 \\
& Na\"{i}ve Bayes     & 82.21 & 88.00 & 82.00 & 79.90 \\
& SVM                 & 88.21 & 89.00 & 88.00 & 88.60 \\
\cline{2-6}
& Zero-Shot (GPT-4o)           & 21.60 & 19.50 & 22.25 & 17.00 \\
& \textbf{Our MA-IDS } & \textbf{90.00} & \textbf{90.00} & \textbf{90.00} & \textbf{89.75} \\
& \quad \textit{Improvement (vs Zero-Shot)} & \textit{+68.40} & \textit{+70.50} & \textit{+67.75} & \textit{+72.75} \\
\hline
\multicolumn{6}{l}{\textit{NF-ToN-IoT}} \\
\hline
& AdaBoost            & 43.83 & 39.67 & 43.83 & 36.79 \\
& Na\"{i}ve Bayes     & 86.19 & 87.61 & 86.19 & 85.94 \\
& SVM                 & 93.79 & 94.66 & 93.79 & 93.42 \\
\cline{2-6}
& Zero-Shot (GPT-4o)           & 13.33 & 11.52 & 11.78 &  4.96 \\
& \textbf{Our MA-IDS } & \textbf{84.00} & \textbf{85.56} & \textbf{85.00} & \textbf{85.22} \\
& \quad \textit{Improvement (vs Zero-Shot)} & \textit{+70.67} & \textit{+74.04} & \textit{+73.22} & \textit{+80.26} \\
\hline
\multicolumn{6}{l}{} \\
\end{tabular}
\end{table*}

\subsection{Dataset and Evaluation Protocol}
We evaluate MA-IDS on two widely used benchmark datasets for IoT intrusion detection, both derived from NetFlow V3 and curated by the University of Queensland~\cite{dataset}. The NF-BoT-IoT dataset comprises over 16.9 million labeled flows, from which we consider four classes: Benign, DDoS, DoS, and Reconnaissance. The NF-ToN-IoT dataset contains 27.5 million flows, from which we evaluate nine classes: Benign, Scanning, DDoS, Backdoor, DoS, Injection, Password, XSS, and MITM.

To ensure a balanced and unbiased evaluation, as well as equitable rule generation across classes, uniform random sampling with equal class quotas is employed. For each dataset, 50,000 samples are used during the library construction phase, while a disjoint set of 20,000 samples is reserved for evaluation. Both subsets are uniformly distributed across all classes, ensuring equitable representation and enabling direct comparison of per-class performance metrics while preventing data leakage between phases.

From the original 53 NetFlow V3 attributes, 14 features are selected based on discriminative relevance, computational efficiency, and privacy preservation. These include contextual identifiers (IP addresses, destination port, protocol), volumetric statistics (byte and packet counts), temporal features (flow duration and inter-arrival times), throughput measures, and TCP flag aggregates, which together capture structural, behavioral, and directional characteristics of network flows. 

Remaining features are excluded for three reasons: (i) redundancy, as many attributes (e.g., min/max packet sizes, TTL, and window statistics) provide overlapping information with selected volumetric and temporal metrics; (ii) limited discriminative contribution, particularly for protocol-specific or rarely populated fields (e.g., DNS, FTP, ICMP); and (iii) incompatibility with privacy-preserving and encrypted traffic settings, where payload-dependent or fine-grained inspection features are unreliable or unavailable.

During preprocessing, invalid values are normalized, protocol identifiers are mapped to categorical representations, and all features are serialized into structured JSON format to enable efficient and interpretable LLM-based reasoning.

\subsection{Evaluation Metrics}
Performance for all configurations was evaluated using overall accuracy along with macro averaged precision, recall, and F1 score. The macro averaging technique calculates each metric independently for every individual class before computing the arithmetic mean across the entire set of classes. This approach assigns equal weight to each traffic category irrespective of the specific sample distribution, which aligns with the uniform sampling strategy employed in this study. 

\begin{figure*}[t]
    \centering
    \begin{subfigure}{0.45\textwidth}
        \centering
        \includegraphics[width=\linewidth]{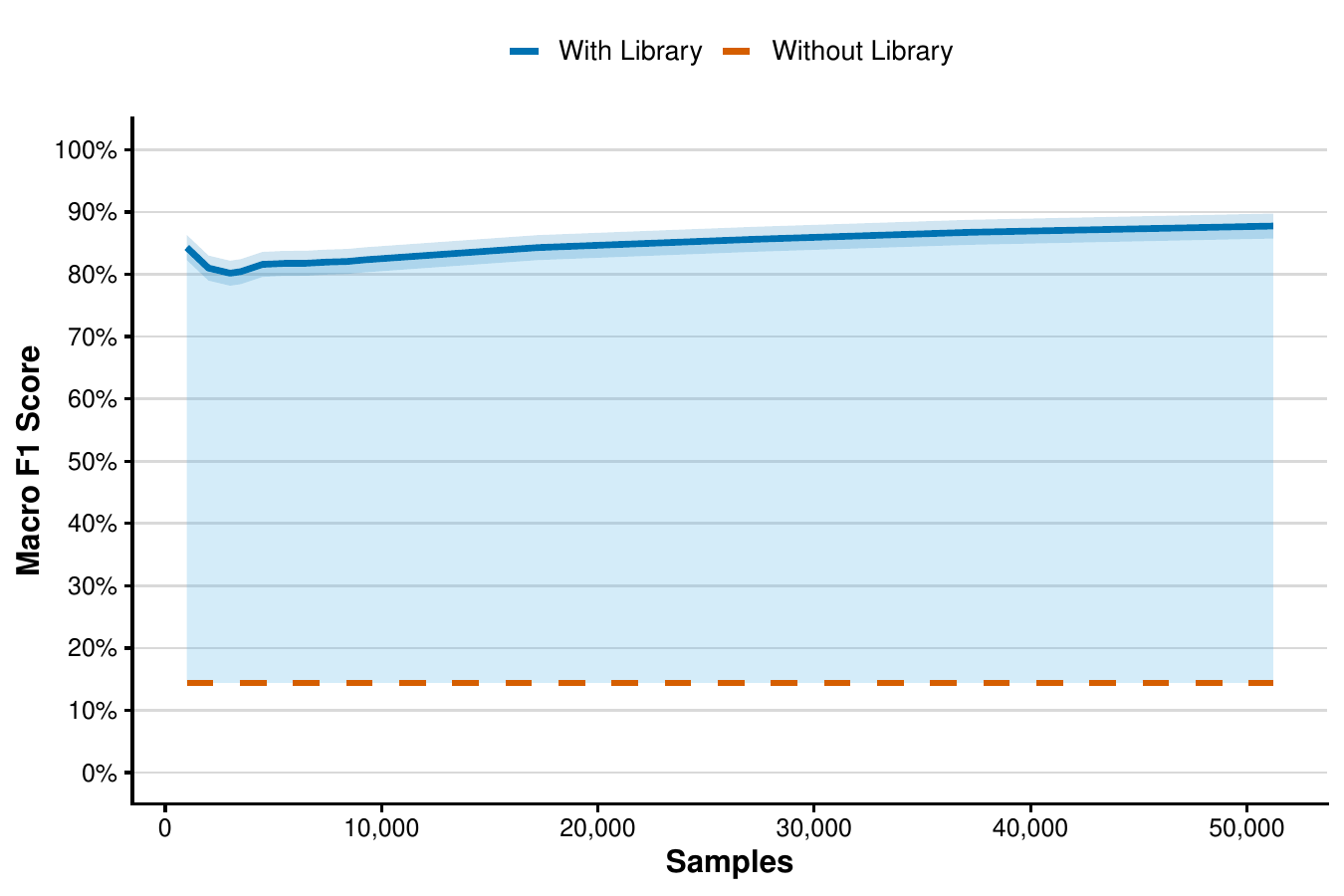}
        \caption{NF-BoT-IoT dataset}
        \label{fig:bot-memory-impact}
    \end{subfigure}
    \hfill
    \begin{subfigure}{0.45\textwidth}
        \centering
        \includegraphics[width=\linewidth]{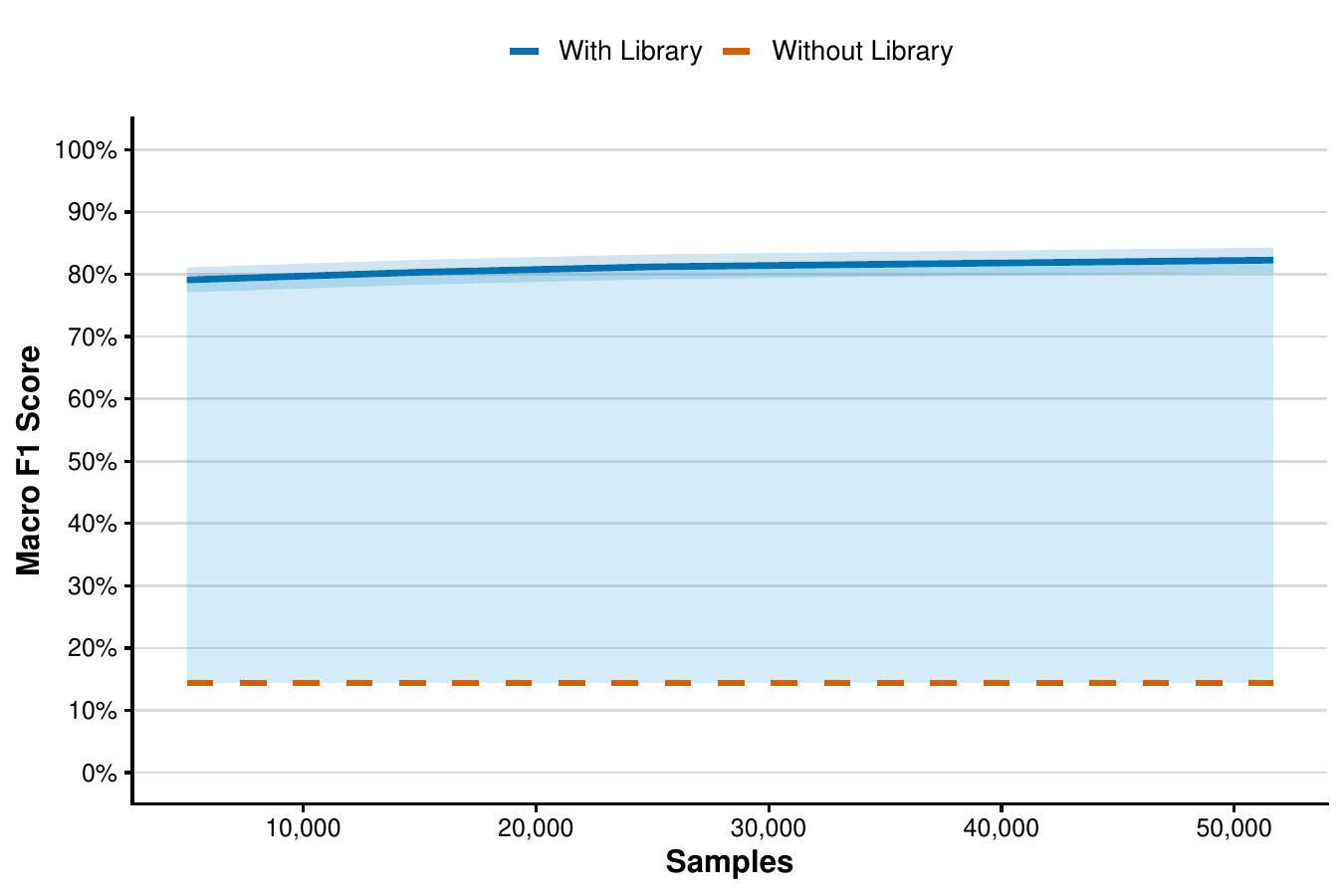}
        \caption{NF-ToN-IoT dataset}
        \label{fig:ton-memory-impact}
    \end{subfigure}
    

    \caption{Macro F1-score over cumulative samples during library construction for MA-IDS with and without the Experience Library on (a) NF-BoT-IoT and (b) NF-ToN-IoT. Performance improves with library growth, while the no-library baseline remains flat, highlighting the impact of accumulated rule context.}
    
    \label{fig:memory-comparison-main}
\end{figure*}

\subsection{Performance Evaluation: NF-BoT-IoT}

The performance of MA-IDS on the NF-BoT-IoT dataset was assessed through a
two-stage experimental protocol. During the library construction stage, the
complete MA-IDS pipeline with the Error Analysis Agent active was executed over
50,000 samples, yielding an accuracy of 85.75\%, Macro Precision of 85.66\%,
Macro Recall of 85.41\%, and Macro F1-Score of 85.43\%. In the evaluation stage,
the Error Analysis Agent was deactivated and the fixed repository of 7,322 induced
rules was used to classify an independent test set of 20,000 samples, achieving
an accuracy of 90.00\% and Macro F1-Score of 89.75\% across all metrics,
demonstrating strong generalisation of the constructed library without further
adaptation (Fig.~\ref{fig:radar-bot}). As reported in Table~\ref{tab:comprehensive-comparison}, MA-IDS at 89.75\% Macro
F1 surpasses SVM (88.60\%), Na\"{i}ve Bayes (79.90\%), and AdaBoost (88.60\%)
on this dataset, while the zero-shot GPT-4o baseline achieves only 17.00\%,
confirming that the Experience Library is the decisive performance driver rather
than the underlying model capability.

\subsection{Performance Evaluation: NF-ToN-IoT}

On the NF-ToN-IoT dataset, the library construction stage over 50,000 samples
with the MA-IDS achieved an accuracy of 80.60\%, Macro
Precision of 83.76\%, Macro Recall of 79.91\%, and a Macro F1-Score of 81.01\%.
In the evaluation stage, with the MA-IDS disabled and the pre-built
library of 9,498 rules frozen, testing on 20,000 samples yielded an accuracy of
84.00\%, Macro Precision of 85.56\%, Macro Recall of 85.00\%, and a Macro
F1-Score of 85.22\%, with the performance contrast between the two configurations
clearly visible in Fig.~\ref{fig:radar-ton}. As reported in Table~\ref{tab:comprehensive-comparison}, MA-IDS at 85.22\% Macro
F1 outperforms AdaBoost (36.79\%) and remains competitive with Na\"{i}ve Bayes
(85.94\%) on this significantly more complex nine-class task, while the zero-shot
GPT-4o baseline collapses to only 4.96\% Macro F1. Although SVM achieves a higher
93.42\%, it offers no interpretability, requires complete retraining. These results confirm that the Experience Library is the critical
enabler of reliable LLM-based classification across both datasets.

\subsection{Ablation Study}

To isolate component contributions, we evaluate three configurations: (1) Zero-Shot Baseline (no retrieval), (2) Library Only (retrieval with fixed Experience Library), and (3) Full MA-IDS (retrieval with continuous rule induction).

On NF-BoT-IoT, the zero-shot baseline achieves 17.00\% Macro F1, with DDoS detection failing (0\% true positives), indicating limited discrimination without contextual grounding. The Library Only configuration improves performance to 89.75\% Macro F1 on 20,000 test samples, demonstrating that retrieval is the primary driver. The Full MA-IDS system attains 85.43\% during the construction phase, where the rule base is still evolving.

On NF-ToN-IoT, the zero-shot baseline degrades to 4.96\% Macro F1, while Library Only reaches 85.22\%, again highlighting the impact of retrieval. Full MA-IDS achieves 81.01\%, with slightly lower performance due to ongoing rule induction. The larger gains on NF-ToN-IoT suggest that retrieval becomes increasingly beneficial as task complexity grows.

Learning curves (Fig.~\ref{fig:memory-comparison-main}) show consistent improvement as the Experience Library expands, while the zero-shot baseline remains unchanged. Performance stabilizes earlier on NF-BoT-IoT and more gradually on NF-ToN-IoT, reflecting differences in class complexity. Overall, retrieval provides the dominant performance gain, while the MA-IDS supports continual adaptation through rule expansion.



\subsection{Experience Library Rule Distribution}
\label{sec:brain-rules}

Table~\ref{tab:brain-rules} summarizes the number of the rules generated per class during the library construction phase. Rule counts reflect class-wise misclassification frequency, with overlapping traffic patterns producing more rules. On NF-BoT-IoT, DDoS and DoS generate the highest counts, while on NF-ToN-IoT, Injection and XSS dominate due to greater semantic complexity. In contrast, Backdoor yields few rules, indicating clear separability. Overall, the library comprises 7,322 rules for NF-BoT-IoT and 9,498 for NF-ToN-IoT, forming the fixed knowledge base for subsequent evaluation.

\begin{table}[htbp]
\centering
\caption{Experience Library Rule Distribution}
\label{tab:brain-rules}
\small
\setlength{\tabcolsep}{10pt}
\renewcommand{\arraystretch}{1.00}
\begin{tabular}{lr | lr}
\hline\hline
\multicolumn{2}{c|}{\textbf{NF BoT IoT}} & \multicolumn{2}{c}{\textbf{NF ToN IoT}} \\
\hline
Class & Rules & Class & Rules \\
\hline
Benign          & 1,631 & Benign    & 1,107 \\
DDoS            & 2,384 & Scanning  & 1,250 \\
DoS             & 1,679 & DDoS      & 708   \\
Reconnaissance  & 1,624 & Backdoor  & 69    \\
Noise           & 4     & DoS       & 745   \\
                &       & Injection & 2,010 \\
                &       & Password  & 820   \\
                &       & XSS       & 1,639 \\
                &       & MITM      & 1,150 \\
\hline
Total  & 7,322 & Total & 9,498 \\
\hline\hline
\end{tabular}
\end{table}

\section{CONCLUSIONS}
This paper presented MA-IDS, a Multi-Agent Intrusion Detection System integrating LLM with Context-Grounded Classification for reasoning-driven intrusion detection in IoT networks. The framework couples a Traffic Classification Agent with an MA-IDS over a persistent Experience Library, enabling
continual self-improvement without model retraining. Evaluation on NF-BoT-IoT and NF-ToN-IoT confirms that the RAG-based Experience
Library is the critical performance driver. Without retrieved context, GPT-4o achieves only 17\% and 4.96\% macro F1 respectively. With Experience Library active, MA-IDS reaches 89.75\% and 85.22\%, accumulating 7,322 and 9,498 detection rules through error-driven induction, competitive with SVM while
offering superior interpretability and adaptability. Future work will investigate lightweight deployment on resource-constrained IoT
devices, extension to open-set zero-day detection, and integration of additional network telemetry to further enrich the Experience Library.





\end{document}